\documentclass[preprint,graphicx,superscriptaddress,longbibliography,onecolumn]{revtex4}
\usepackage{graphicx,amsfonts}
\usepackage{amsmath}

\begin{document}


\title{High-charge relativistic electrons by vacuum laser acceleration from plasma mirrors using flying focus pulses} 



\author{Jiaxin Liu}
\affiliation{Key Laboratory of High Energy Density Physics and Technology (MoE), College of Physics, Sichuan University, Chengdu 610064, China}
\author{Zeyue Pang}
\affiliation{Key Laboratory of High Energy Density Physics and Technology (MoE), College of Physics, Sichuan University, Chengdu 610064, China}
\author{Hehanlin Wang}
\affiliation{Key Laboratory of High Energy Density Physics and Technology (MoE), College of Physics, Sichuan University, Chengdu 610064, China}
\author{Zi-Yu Chen}
\email[]{ziyuch@scu.edu.cn}
\affiliation{Key Laboratory of High Energy Density Physics and Technology (MoE), College of Physics, Sichuan University, Chengdu 610064, China}

%


\date{\today}

\begin{abstract}
Relativistic electron beams produced by intense lasers over short distances have important applications in high energy density physics and medical technologies. Vacuum laser acceleration with plasma mirrors injectors has garnered substantial research interest recently. However, a persistent challenge remains unresolved that electrons inevitably detach from the laser acceleration phase due to velocity mismatch. Here, we employ flying focus lasers to address this limitation. Through three-dimensional particle-in-cell simulations, we demonstrate that flying focus lasers can achieve a substantial enhancement in relativistic electron charge yield compared to conventional Gaussian lasers. This improvement stems from two key attributes: (1) The subluminal propagation velocity of the peak intensity keeps a larger electron population synchronized within the longitudinal ponderomotive acceleration region, and (2) Flying focus lasers sustain higher magnitudes of the longitudinal ponderomotive force over longer distances in comparison to Gaussian lasers. This approach offers high-charge relativistic electron sources ideal for demanding applications such as high-flux Thomson scattering and radiography.
\end{abstract}


\maketitle 

\section{Introduction}
The development of ultraintense femtosecond lasers \cite{yu2012generation,yoon2021realization} has enabled electron acceleration to relativistic energies over very short distance \cite{esarey2009physics}, paving the way for compact accelerator technologies with broad applications including ultrafast electron diffraction \cite{morimoto2018diffraction,he2016capturing}, fast ignition \cite{tabak2005review} and radiation therapy \cite{fuchs2009treatment}. Over the past decades, two primary schemes have been extensively investigated both theoretically and experimentally: (1) Laser wakefield acceleration, in which intense lasers drive plasma wakefields in underdense plasmas, creating accelerating gradients typically on the order of 100 GV/m that can energize electrons to several GeV \cite{tajima1979laser,gonsalves2019petawatt}; and (2) Vacuum laser acceleration (VLA), where electrons are directly accelerated by the intense fields provided by the laser pulses with gradients exceeding 10 TV/m \cite{esarey1995laser,malka1997experimental,varin2006relativistic,de2024unforeseen}. 

In VLA, electron acceleration occurs in a vacuum environment where plasma field effects are negligible. This simplified mechanism has attracted significant efforts to enhance the accelerated beam qualities. However, electron injection in VLA is a critical challenge. For effective energy transfer from the laser field to electrons, two key prerequisites must be met: (1) an optimal injection position to align with the accelerating phase of the laser field, and (2) an initial injection velocity close to the speed of light to ensure sustained synchronization with the laser pulse. The recent concept of employing relativistic plasma mirrors as injectors offers an effective solution to both requirements \cite{thevenet2016vacuum}. When a $ p $-polarized ultraintense laser obliquely irradiates a solid plasma mirror, surface electrons are periodically ejected at relativistic speeds via the relativistic oscillating mirror mechanism \cite{bulanov1994interaction,lichters1996short,baeva2006theory,thaury2007plasma}, naturally accumulating near the zero-crossing of the reflected laser's electric field. Relativistic electron bunches with central energy reaching 10 MeV have been observed from plasma mirrors using 20-TW linearly polarized lasers \cite{thevenet2016vacuum}. Building upon this, subsequent studies have explored VLA using tightly focused radially polarized laser pulses, which generate a strong longitudinal electric field to drive acceleration \cite{PhysRevLett.119.094801,zaim2020interaction,cao2021direct}. To further enhance both the acceleration process and beam confinement, twisted lasers with a topological charge of $ |l|=1 $, which exhibit both longitudinal electric and magnetic fields, have been proposed for generating monoenergetic attosecond electron bunches \cite{shi2021generation}.

Despite these advancements, a key challenge persists: electrons inevitably detach from the acceleration phase during the interaction. In conventional laser beams, the high-intensity region and peak intensity propagate at the speed of light $ c $ in vacuum. However, since electrons cannot reach this speed, they inevitably fall behind the pulse envelope and exit the acceleration region. Additionally, the inherent divergence of conventional laser pulses beyond their focal waist, constrained by the Rayleigh range, leads to a decrease in peak intensity, thereby limiting efficient acceleration over long propagation distances.

Recently, a novel laser technique known as the flying focus laser pulse has been developed\cite{froula2019flying}. In this approach, the focal velocity can be decoupled from the group velocity of the pulse envelope, enabling the focus to propagate at any desired speed. Flying focus pulses can be generated either by controlling the focal time of a chirped pulse using a diffractive lens or by manipulating the focal position of different segments of an unchirped laser pulse through a radially dependent spatiotemporal coupling\cite{jolly2020controlling,froula2018spatiotemporal,sainte2017controlling,smartsev2019axiparabola,simpson2022spatiotemporal,ambat2023programmable,pigeon2023ultrabroadband,PhysRevResearch.5.013085}. This technique has been proposed as a promising solution to the dephasing problem encountered in laser wakefield acceleration\cite{palastro_dephasingless_2020,caizergues_phase-locked_2020}. Additionally, flying focus pulses endowed with orbital angular momentum have been shown to provide transverse confinement of ultrarelativistic charged particle bunches over macroscopic distances\cite{formanek_charged_2023}.

In this paper, we propose VLA from plasma mirrors utilizing flying focus lasers to address the limitations of conventional lasers. By setting the focal velocity to subluminal values, high-energy electrons can remain near the rising edge of the pulse envelope, within the longitudinal ponderomotive acceleration zone \cite{ramsey2020vacuum,ramsey2022nonlinear}, thereby staying in phase with the laser's acceleration process. Moreover, flying focus beams can sustain near-constant peak intensities over longer distances compared to traditional Gaussian lasers, thus overcoming Rayleigh range limitations. These features can work synergistically to enhance the quality of electron beams in VLA. Using three-dimensional (3D) particle-in-cell (PIC) simulations, we demonstrate that flying focus lasers achieve a substantial increase, up to an order of magnitude, in the charge of relativistic electron bunches compared to conventional laser pulses.

\section{Simulation setup}

\begin{figure}[ht!]
	\centering
	\includegraphics[width=0.9\linewidth]{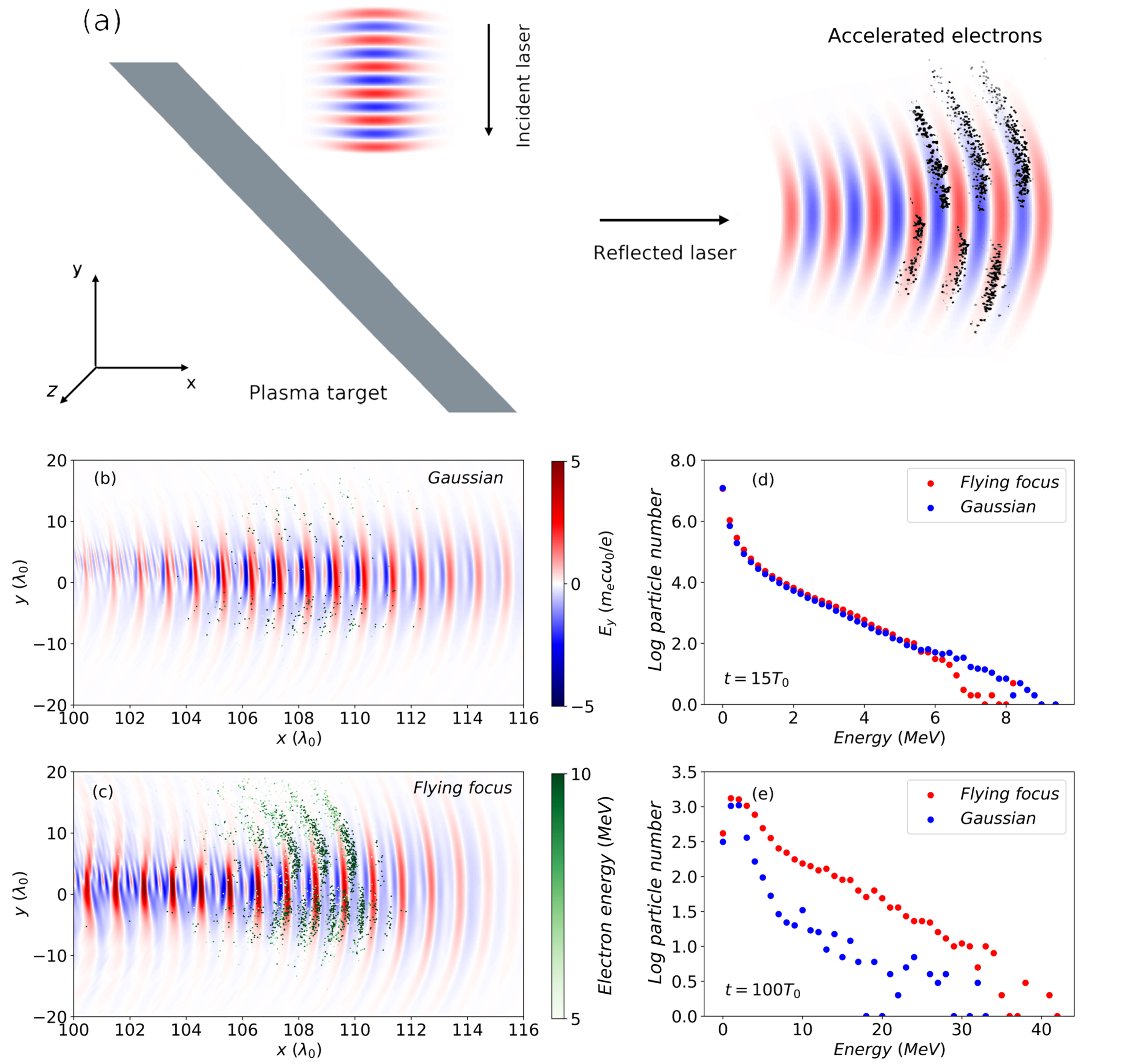}
	\caption{(a) Schematic illustration of the simulation setup. A relativistic laser pulse with $ a_0=3 $, incident at a $45^{\circ}$ angle along the $ y $-axis, irradiates a solid-density plasma target. Upon reflection, the electrons injected from the plasma surface are accelerated by the reflected laser fields propagating along the $ x $-axis. (b)-(c) Snapshots of the electron bunches ($\ge5$ MeV) distribution at $t=100T_{0}$ for the (b) Gaussian and (c) flying focus lasers. The electrons are represented by green scattered dots, while the distribution of the electric field $E_y$ along the polarization direction is shown in a blue-red color scale. (d)-(e) Comparison of the electron energy spectra generated by the Gaussian (blue dots) and flying focus lasers (red dots) at (d) $t=15T_{0}$ (d) and (e) $t=100T_{0}$.}
	\label{figure1}
\end{figure}

The 3D PIC simulations are performed using the VLPL (Virtual Laser Plasma Lab) code \cite{pukhov1999three}. An $ x $-polarized flying focus beam is incident along the $ -y $ direction and focused onto an overdense plasma target at a 45$^{\circ}$ incidence angle. The reflected laser pulses propagate and accelerate the electrons injected from the plasma mirror along the $ +x $ direction [see Fig. \ref{figure1}(a)]. The initial electric field of the flying focus pulse can be written as \cite{ramsey2022nonlinear,ramsey2023exact}:
\begin{equation}\label{ff_expr}
	\begin{split}
		E=E_{0} \frac{w_{0}}{w(\xi^{\prime})} \exp \Big[i(k_{0}y-\omega_{0}t)-\Big(1-i \frac{\xi^{\prime}}{\xi_{0}^{\prime}}\Big)
		\frac{r^{2}}{w^{2}(\xi^{\prime})}-i\arctan\Big(\frac{\xi^{\prime}}{\xi_{0}^{\prime}}\Big)\Big]T(t),
	\end{split}
\end{equation}
where $E_0=a_{0}m_{e}\omega_{0}c/e$ is the electric field amplitude at focus with the normalized laser vector potential $a_0=3$, $m_e$ is the electron rest mass, $c$ is the speed of light in vacuum, $e$ is the elementary charge, $w(\xi^{\prime})=w_{0}\sqrt{1+(\xi^{\prime}/\xi_{0}^{\prime})^{2}}$ is the beam waist, $w_0=5\lambda_0$ is the minimum beam waist, $\lambda_0=800$ nm is the laser wavelength, $\xi^{\prime}=y-\upsilon_{f}t$ with $\upsilon_{f}=0.95c$ being the velocity of focus, $\xi_{0}^{\prime}=\lvert 1-\upsilon_{f}/c \rvert Z_{R}$ with $Z_{R}=k_{0}w_{0}^{2}/2$ being the Rayleigh length, $k_0$ and $\omega_0$ are respectively the laser wave number and angular frequency, $r=\sqrt{x^{2}+z^{2}}$ is the radial distance, $T(t)=\exp[-(t^{2}/2\tau^{2})^{8}]$ is a longitudinal super-Gaussian envelope as employed in \cite{franke_optical_2021} to ensure the peak intensity of the flying focus beam maintain nearly constant during focal spot translation \cite{ramsey2022nonlinear}. To guarantee a near-constant maximum intensity even for a focus velocity of $0.9c$, we set a relatively long temporal envelope with $ \tau=17T_0 $, where $ T_0 $ is the laser period.

The fully ionized plasma has a density of $n_0 = 100n_c$ and a thickness of 0.2$\lambda_{0}$, where $n_c = m_{e}\omega_{0}^{2}/4\pi e^{2}$ is the critical density. The density decays exponentially with a scale length of $0.1\lambda_0$ at the front of the target. The ions are assumed to be immobile in the simulations\cite{shi2021generation}.
The simulation box has a size of $X \times Y \times Z = 36\lambda_0 \times 40\lambda_0 \times 40\lambda_0$. The grid step size in each direction is $0.032\lambda_0$ and the time step is $0.018T_0$. Absorption boundary condition is applied to both particles and fields. Limited by the significant computational time and storage requirements of 3D simulations, each cell is filled with 1 macroparticles. We have performed convergence tests, which show the key results, such as the accelerated charge, remaining stable as the number of macroparticles per cell is increased. To track the electron dynamics far from the target, a moving window technique is employed, advancing at a speed of 0.9$c$ along the +$ x $ direction.

\section{Results}

Figures \ref{figure1}(b) and \ref{figure1}(c) show the spatial distributions of the electric fields $ E_y $ and the accelerated electrons with energy $\ge5$ MeV, projected onto the $z=0$ plane, generated by conventional Gaussian and flying focus laser pulses at $t=100T_{0}$, respectively. Both lasers arrive at the target surface simultaneously, with identical peak intensities ($ a_0=3 $) and full width at half maximum (FWHM) pulse durations of $ 11T_0 $. The FWHM pulse duration for the flying focus pulse is determined numerically by calculating the FWHM of the pulse’s temporal profile as directly observed in the PIC simulation results. We then set the FWHM of the Gaussian pulse to the same value. This is done to ensure that both pulses have approximately the same waveform in their leading edges, thereby producing a similar ponderomotive force for longitudinal acceleration. 

A train of femtosecond relativistic electron bunches is produced through VLA in both cases. However, a striking feature is that the number of electrons undergoing efficient acceleration and remaining synchronized with the laser field is significantly higher for flying focus lasers compared to Gaussian lasers. Additionally, the dephasing effect between the electrons and the laser field is evident, as some electrons are observed surfing in the deceleration phase. 

Figure \ref{figure1}(d) compares the energy spectra of all electrons at $t=15T_{0}$, when the laser peak first arrives at the plasma surface. It shows that the initial electron energy spectra generated by the two lasers are very similar. In fact, the high-energy tail of electrons produced by the Gaussian laser is even more pronounced than that from the flying focus laser. 
After the VLA process, however, the electron energy spectra at $t=100T_{0}$, shown in Fig. \ref{figure1}(e), indicate that the cutoff energy of the accelerated electrons driven by the flying focus lasers is now higher than that of the Gaussian lasers. More importantly, the number of relativistic electrons driven by the flying focus lasers is significantly greater, approximately one order of magnitude higher, than that produced by the Gaussian lasers. Specifically, the total charge of the accelerated electrons with energy $\ge5$ MeV reaches 1.4 nC for the flying focus laser case, compared to only 0.2 nC for the Gaussian laser case.

As we set a relatively long temporal envelope for the flying focus pulse, the resulting waveform is asymmetric with a very long tail. Consequently, the energy of the flying focus pulse is 2.6 times that of the Gaussian pulse. Specifically, the total energy in the Gaussian and flying focus laser pulse is approximately 100 mJ and 260 mJ, respectively. Nevertheless, the charge of the accelerated electrons from the flying focus pulse is up to 7 times greater than that of the Gaussian pulse. Crucially, the accelerated electrons are located in the front part of the pulse, meaning the long tail of the flying focus pulse does not contribute to electron acceleration. This excessively long pulse portion and the corresponding extra energy are actually unnecessary. It is foreseeable that by shortening the temporal envelope width of the flying focus pulse to be comparable to that of the Gaussian pulse, the enhancement of accelerated electron charge by the flying focus pulses would remain approximately the same as the current result.

\begin{figure}[ht!]
	\centering
	\includegraphics[width=0.9\linewidth]{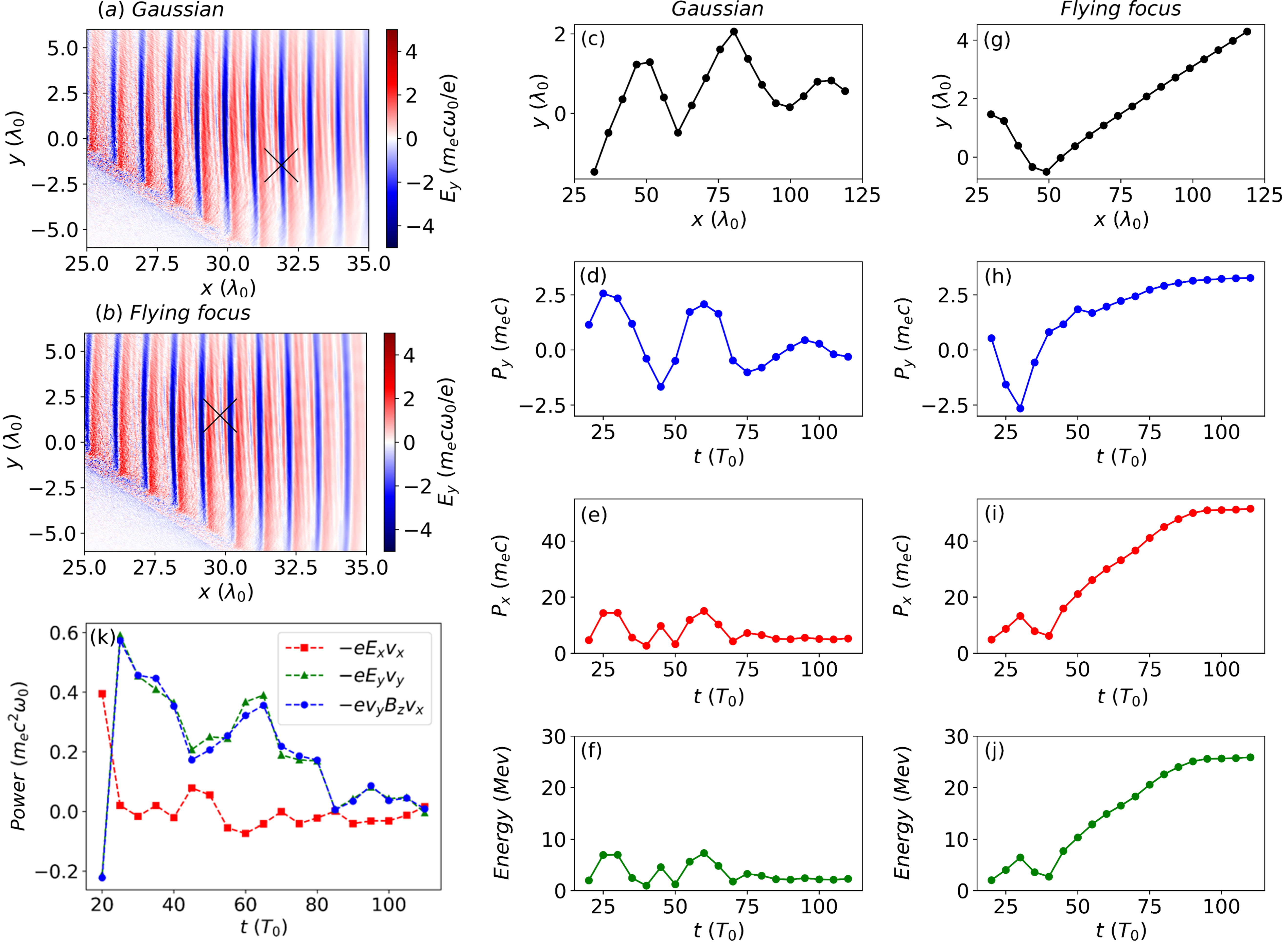}
	\caption{(a)--(b) Spatial distribution of representative electrons within the electric field $E_{y}$ driven by the (a) Gaussian laser and (b) flying focus laser, respectively, at the moment of electron emission from the target surface at $t=20T_{0}$. (c)--(f) Dynamics of the electron from panel (a), showing: (c) its trajectory in the x-y plane; temporal evolution of (d) transverse momentum, (e) longitudinal momentum, and (f) total energy. (g)--(j) These subfigures correspond to (c)--(f) for the electron shown in panel (b). (k) The instantaneous power delivered to a single electron by each field component as a function of time.}
	\label{figure2}
\end{figure}

To understand the differences in acceleration outcomes, we trace the evolution of electron dynamics during VLA. Figures \ref{figure2}(a) and \ref{figure2}(b) illustrate the initial positions of representative electrons shortly after their injection from plasma mirrors into the $ E_y $ electric fields of Gaussian and flying focus lasers, respectively. Notably, the electron within the Gaussian laser field is in the acceleration phase, while the electron within the flying focus laser field is being decelerated. 
Figure \ref{figure2}(c) shows the electron's $ x $-$ y $ plane trajectory within the Gaussian laser field. The electron undergoes predominantly longitudinal acceleration for approximately 100$ \lambda_{0} $, accompanied by transverse oscillations. However, the final energy gain remains low. Figures \ref{figure2}(d)--\ref{figure2}(f) display the temporal evolution of its transverse momentum $ p_y $, longitudinal momentum $ p_x $, and energy, respectively, all exhibiting oscillatory behavior. These oscillations, indicative of interaction with multiple Gaussian laser field optical cycles, arises from a lack of phase synchronization. The insufficient longitudinal momentum growth [Fig. \ref{figure2}(f)] prevents the electron from maintaining phase with the laser pulse, leading to repeated acceleration and deceleration, and ultimately restricting the net energy transfer. In contrast, the electron, despite initial deceleration in the flying focus laser field, shows a monotonic rise in momenta and energy after a single oscillation [see Figs. \ref{figure2}(g)--\ref{figure2}(j)]. It is noteworthy that both multi-cycle interacting and phase-locked electrons are present in VLA for both Gaussian and flying focus beams. However, the crucial distinction is the significantly enhanced yield of phase-locked electrons in the flying focus beam's VLA, driving its enhanced relativistic electron beam production. 
For the flying focus pulse, 0.88\% of the initially injected electrons ($>0.5$ MeV at $ t=15T_0 $) remain in the accelerating phase after a considerable amount of time ($>5$ MeV at $ t=100T_0 $). In contrast, this percentage is only 0.16\% for the Gaussian pulse.

As depicted in Fig.~\ref{figure2}(g), although the trajectory of the accelerated electron exhibits a continuous drift along the polarization direction, it remains spatially confined within the laser spot region. The observed increase in longitudinal momentum [Fig.~\ref{figure2}(i)] is substantially greater than that of the transverse momentum [Fig.~\ref{figure2}(h)], signifying that the electron experiences primarily longitudinal acceleration. Figure \ref{figure2}(k) shows the instantaneous work done by each field component as a function of time. It is evident that the longitudinal acceleration is predominantly driven by the combined contribution of the transverse magnetic $B_{z}$ and electric $E_{y}$ fields, highlighting the critical role of the laser’s longitudinal ponderomotive force. 

\begin{figure}[ht!]
	\centering
	\includegraphics[width=0.9\linewidth]{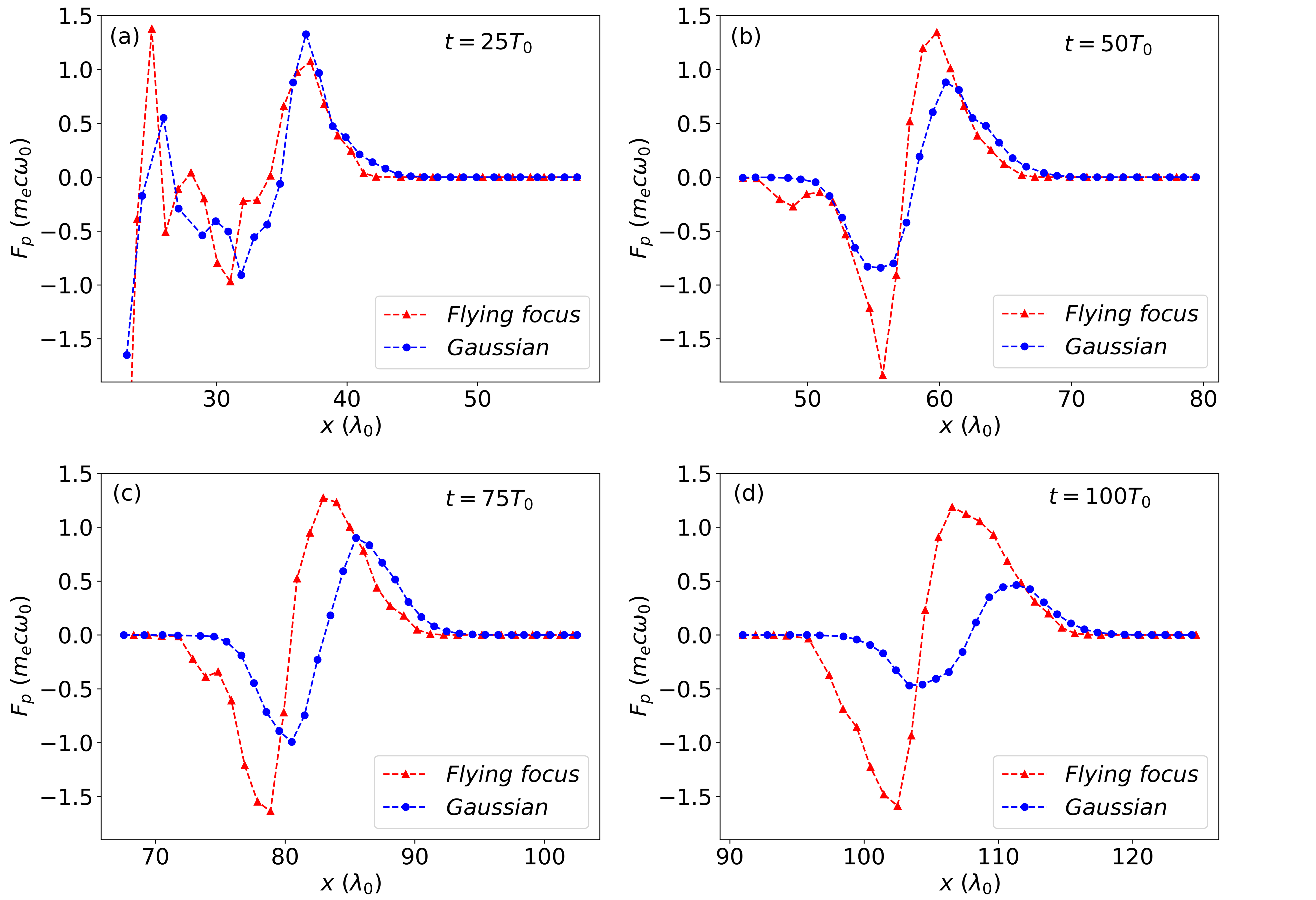}
	\caption{The spatial profiles of longitudinal ponderomotive forces derived from the on-axis intensities of the Gaussian laser (red markers) and flying focus laser (blue markers), respectively, after reflection from the plasma target at different time instants: (a) $t=25T_{0}$, (b) $t=50T_{0}$, (c) $t=75T_{0}$, (d) $t=100T_{0}$. The longitudinal ponderomotive force is calculated as $F_{p}=-e^{2}\nabla_{x}E_{y}^{2}\ /\ 4m_{e}\omega_{0}^{2}$, where $ E_y^2 $ is sampled at intervals of one optical cycle.}
	\label{figure3}
\end{figure}

We then examine the spatial profile evolution of the laser's longitudinal ponderomotive force at various acceleration stages (Figure \ref{figure3}). Given the identical transverse Gaussian profiles of both flying focus and Gaussian lasers, as can be seen from Eq. (\ref{ff_expr}), we analyze the one-dimensional longitudinal ponderomotive forces along the optical axis. Initially, the peak strength of the two lasers' longitudinal ponderomotive force are comparable. However, the peak intensity of Gaussian lasers propagates at the speed of light $ c $, inevitably causing electrons to move out of the ponderomotive acceleration region and into the deceleration region. The phase-mismatched electrons interact with multiple laser periods, undergoing repeated cycles of acceleration and deceleration, which ultimately results in low energy gain. Additionally, due to diffraction over the Rayleigh length (approximately 78$ \lambda_{0} $), the peak strength of the Gaussian lasers' ponderomotive force decreases considerably. In contrast, the flying focus lasers maintain the peak strength of their longitudinal ponderomotive force over much longer distances, resulting in a higher cutoff energy for the accelerated electrons. Moreover, the peak intensity of the flying focus lasers propagates at a preset subluminal velocity of $v=0.95c$. As a result, a large population of high-energy electrons can remain within the ponderomotive acceleration region throughout the interaction, leading to a significantly larger number of accelerated electrons. The ponderomotive force for longitudinal electron acceleration is determined by the final generated pulse shape, which is a result of the combined effect of the Rayleigh range prefactor and the temporal envelope function. We set the parameters so that the front half-waveforms of the two pulses are approximately the same, ensuring that the initial ponderomotive force in the accelerating part is similar. The subluminal focal velocities and the sustained high ponderomotive force of flying focus pulses work synergistically to deliver exceptional acceleration performance.

\begin{figure}[ht!]
	\centering
	\includegraphics[width=0.9\linewidth]{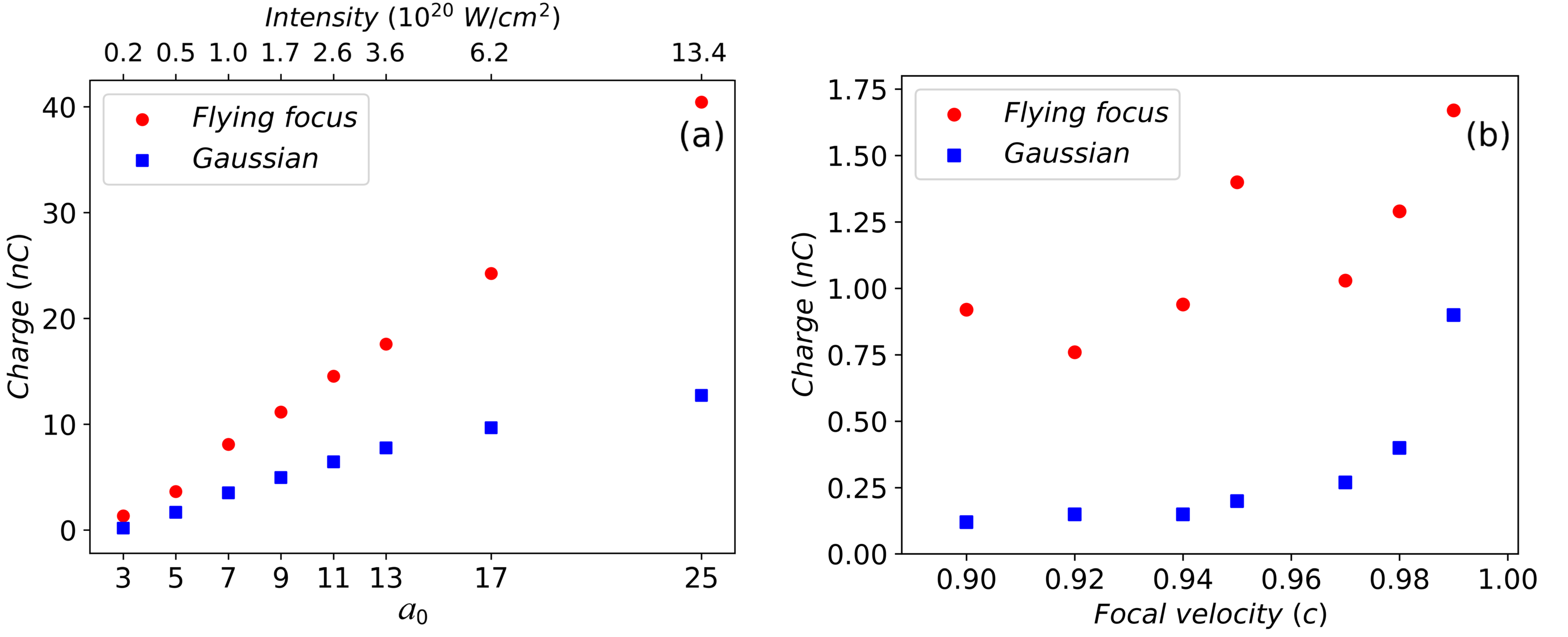}
	\caption{The charge of high-energy electron bunches (with energies $\ge5$ MeV) generated by the Gaussian and flying focus lasers as a function of (a) the laser dimensionless parameter $a_{0}$ and intensity, and (b) the focal velocity. The energy of flying focus pulses in (a) spans from approximately 268 mJ up to 2.2 J. The FWHM pulse durations of the flying focus and corresponding Gaussian pulses in (b) are $ 14T_0 $, $ 13T_0 $, $ 12T_0 $, $ 11T_0 $, $ 8T_0 $, $ 6T_0 $, and $ 5T_0 $ at focal velocities of $0.90 c$, $0.92 c$, $0.94 c$, $0.95 c$, $0.97 c$, $0.98 c$, and $0.99 c$, respectively.}
	\label{figure4}
\end{figure}

Flying focus lasers can significantly enhance VLA electron charges across a broad range of laser-plasma parameters. Figure \ref{figure4}(a) shows the charge of accelerated electrons with energy $\ge5$ MeV as a function of laser intensity. While Gaussian lasers also exhibit increased electron charge with intensity due to enhanced ponderomotive force and injection velocities, the flying focus laser demonstrates a steeper rise in electron charge. At an intensity level of approximately $ 10^{20} $ W/cm$ ^{2} $, flying focus lasers achieve a total charge of approximately 10 nC for electrons with energies $\ge5$ MeV. Such ultrashort relativistic electron bunches with substantial charge are highly advantageous for various applications, including single-shot electron radiography and the generation of high-flux x-rays \cite{courtois2011high,bruhaug2023single}. 

In addition to laser peak intensity, the focal velocity of the flying focus lasers also has an impact on the results. However, the expression for the flying focus beam employed in this study [Eq. (\ref{ff_expr})] inherently couples focal velocity adjustments to changes in pulse duration, and its variation influences the ponderomotive force, Rayleigh range, and accelerating phase length. This coupling makes it challenging to conduct an isolated study of the effects of focal velocity while maintaining a constant ponderomotive force. Nevertheless, we perform simulations with different focal velocity and pulse width. We have also found significant enhancement in the charges of accelerated electrons for the flying focus pulse compared to the Gaussian pulse with the same FWHM pulse duration for varying focal velocities [Fig. \ref{figure4}(b)]. For the flying focus pulse, increasing the focal velocity reduces the pulse width, which increases the ponderomotive force but simultaneously shrinks the Rayleigh range and possibly the accelerating phase length. Because the final acceleration outcome is a combination of these effects, the total accelerated charge does not consistently increase. In contrast, for the Gaussian pulse, accelerated charge simply increases as the pulse width decreases and the ponderomotive force rises.

\section{Conclusion}
In conclusion, 3D PIC simulations reveal that flying focus lasers dramatically improve the yield of high-energy electrons in the VLA scheme with plasma mirrors. This enhancement is driven by two primary characteristics. First, their subluminal intensity peak propagation allows for sustained synchronization between electrons and the laser field. Second, in comparison to a Gaussian beam, the flying focus beam maintains higher longitudinal ponderomotive force magnitudes over extended temporal and spatial scales. This approach offers superior relativistic electron sources for applications such as single-shot radiography and Thomson scattering, where high-flux beams are essential for achieving more accurate and detailed imaging and measurements.


\section*{acknowledgments}
This work was supported by the National Natural Science Foundation of China (12175157).

\bibliography{ref_ff_vla}

\begin{thebibliography}{41}
\expandafter\ifx\csname natexlab\endcsname\relax\def\natexlab#1{#1}\fi
\expandafter\ifx\csname bibnamefont\endcsname\relax
  \def\bibnamefont#1{#1}\fi
\expandafter\ifx\csname bibfnamefont\endcsname\relax
  \def\bibfnamefont#1{#1}\fi
\expandafter\ifx\csname citenamefont\endcsname\relax
  \def\citenamefont#1{#1}\fi
\expandafter\ifx\csname url\endcsname\relax
  \def\url#1{\texttt{#1}}\fi
\expandafter\ifx\csname urlprefix\endcsname\relax\def\urlprefix{URL }\fi
\providecommand{\bibinfo}[2]{#2}
\providecommand{\eprint}[2][]{\url{#2}}

\bibitem[{\citenamefont{Yu et~al.}(2012)\citenamefont{Yu, Lee, Sung, Yoon,
  Jeong, and Lee}}]{yu2012generation}
\bibinfo{author}{\bibfnamefont{T.~J.} \bibnamefont{Yu}},
  \bibinfo{author}{\bibfnamefont{S.~K.} \bibnamefont{Lee}},
  \bibinfo{author}{\bibfnamefont{J.~H.} \bibnamefont{Sung}},
  \bibinfo{author}{\bibfnamefont{J.~W.} \bibnamefont{Yoon}},
  \bibinfo{author}{\bibfnamefont{T.~M.} \bibnamefont{Jeong}}, \bibnamefont{and}
  \bibinfo{author}{\bibfnamefont{J.}~\bibnamefont{Lee}},
  \bibinfo{journal}{Optics Express} \textbf{\bibinfo{volume}{20}},
  \bibinfo{pages}{10807} (\bibinfo{year}{2012}).

\bibitem[{\citenamefont{Yoon et~al.}(2021)\citenamefont{Yoon, Kim, Choi, Sung,
  Lee, Lee, and Nam}}]{yoon2021realization}
\bibinfo{author}{\bibfnamefont{J.~W.} \bibnamefont{Yoon}},
  \bibinfo{author}{\bibfnamefont{Y.~G.} \bibnamefont{Kim}},
  \bibinfo{author}{\bibfnamefont{I.~W.} \bibnamefont{Choi}},
  \bibinfo{author}{\bibfnamefont{J.~H.} \bibnamefont{Sung}},
  \bibinfo{author}{\bibfnamefont{H.~W.} \bibnamefont{Lee}},
  \bibinfo{author}{\bibfnamefont{S.~K.} \bibnamefont{Lee}}, \bibnamefont{and}
  \bibinfo{author}{\bibfnamefont{C.~H.} \bibnamefont{Nam}},
  \bibinfo{journal}{Optica} \textbf{\bibinfo{volume}{8}}, \bibinfo{pages}{630}
  (\bibinfo{year}{2021}).

\bibitem[{\citenamefont{Esarey et~al.}(2009)\citenamefont{Esarey, Schroeder,
  and Leemans}}]{esarey2009physics}
\bibinfo{author}{\bibfnamefont{E.}~\bibnamefont{Esarey}},
  \bibinfo{author}{\bibfnamefont{C.~B.} \bibnamefont{Schroeder}},
  \bibnamefont{and} \bibinfo{author}{\bibfnamefont{W.~P.}
  \bibnamefont{Leemans}}, \bibinfo{journal}{Reviews of Modern Physics}
  \textbf{\bibinfo{volume}{81}}, \bibinfo{pages}{1229} (\bibinfo{year}{2009}).

\bibitem[{\citenamefont{Morimoto and Baum}(2018)}]{morimoto2018diffraction}
\bibinfo{author}{\bibfnamefont{Y.}~\bibnamefont{Morimoto}} \bibnamefont{and}
  \bibinfo{author}{\bibfnamefont{P.}~\bibnamefont{Baum}},
  \bibinfo{journal}{Nature Physics} \textbf{\bibinfo{volume}{14}},
  \bibinfo{pages}{252} (\bibinfo{year}{2018}).

\bibitem[{\citenamefont{He et~al.}(2016)\citenamefont{He, Beaurepaire, Nees,
  Gall{\'e}, Scott, P{\'e}rez, Lagally, Krushelnick, Thomas, and
  Faure}}]{he2016capturing}
\bibinfo{author}{\bibfnamefont{Z.-H.} \bibnamefont{He}},
  \bibinfo{author}{\bibfnamefont{B.}~\bibnamefont{Beaurepaire}},
  \bibinfo{author}{\bibfnamefont{J.}~\bibnamefont{Nees}},
  \bibinfo{author}{\bibfnamefont{G.}~\bibnamefont{Gall{\'e}}},
  \bibinfo{author}{\bibfnamefont{S.}~\bibnamefont{Scott}},
  \bibinfo{author}{\bibfnamefont{J.~S.} \bibnamefont{P{\'e}rez}},
  \bibinfo{author}{\bibfnamefont{M.}~\bibnamefont{Lagally}},
  \bibinfo{author}{\bibfnamefont{K.}~\bibnamefont{Krushelnick}},
  \bibinfo{author}{\bibfnamefont{A.}~\bibnamefont{Thomas}}, \bibnamefont{and}
  \bibinfo{author}{\bibfnamefont{J.}~\bibnamefont{Faure}},
  \bibinfo{journal}{Sci. Rep.} \textbf{\bibinfo{volume}{6}},
  \bibinfo{pages}{36224} (\bibinfo{year}{2016}).

\bibitem[{\citenamefont{Tabak et~al.}(2005)\citenamefont{Tabak, Clark,
  Hatchett, Key, Lasinski, Snavely, Wilks, Town, Stephens, Campbell
  et~al.}}]{tabak2005review}
\bibinfo{author}{\bibfnamefont{M.}~\bibnamefont{Tabak}},
  \bibinfo{author}{\bibfnamefont{D.}~\bibnamefont{Clark}},
  \bibinfo{author}{\bibfnamefont{S.}~\bibnamefont{Hatchett}},
  \bibinfo{author}{\bibfnamefont{M.}~\bibnamefont{Key}},
  \bibinfo{author}{\bibfnamefont{B.}~\bibnamefont{Lasinski}},
  \bibinfo{author}{\bibfnamefont{R.}~\bibnamefont{Snavely}},
  \bibinfo{author}{\bibfnamefont{S.}~\bibnamefont{Wilks}},
  \bibinfo{author}{\bibfnamefont{R.}~\bibnamefont{Town}},
  \bibinfo{author}{\bibfnamefont{R.}~\bibnamefont{Stephens}},
  \bibinfo{author}{\bibfnamefont{E.}~\bibnamefont{Campbell}},
  \bibnamefont{et~al.}, \bibinfo{journal}{Physics of Plasmas}
  \textbf{\bibinfo{volume}{12}}, \bibinfo{pages}{057305}
  (\bibinfo{year}{2005}).

\bibitem[{\citenamefont{Fuchs et~al.}(2009)\citenamefont{Fuchs, Szymanowski,
  Oelfke, Glinec, Rechatin, Faure, and Malka}}]{fuchs2009treatment}
\bibinfo{author}{\bibfnamefont{T.}~\bibnamefont{Fuchs}},
  \bibinfo{author}{\bibfnamefont{H.}~\bibnamefont{Szymanowski}},
  \bibinfo{author}{\bibfnamefont{U.}~\bibnamefont{Oelfke}},
  \bibinfo{author}{\bibfnamefont{Y.}~\bibnamefont{Glinec}},
  \bibinfo{author}{\bibfnamefont{C.}~\bibnamefont{Rechatin}},
  \bibinfo{author}{\bibfnamefont{J.}~\bibnamefont{Faure}}, \bibnamefont{and}
  \bibinfo{author}{\bibfnamefont{V.}~\bibnamefont{Malka}},
  \bibinfo{journal}{Physics in Medicine \& Biology}
  \textbf{\bibinfo{volume}{54}}, \bibinfo{pages}{3315} (\bibinfo{year}{2009}).

\bibitem[{\citenamefont{Tajima and Dawson}(1979)}]{tajima1979laser}
\bibinfo{author}{\bibfnamefont{T.}~\bibnamefont{Tajima}} \bibnamefont{and}
  \bibinfo{author}{\bibfnamefont{J.~M.} \bibnamefont{Dawson}},
  \bibinfo{journal}{Physical Review Letters} \textbf{\bibinfo{volume}{43}},
  \bibinfo{pages}{267} (\bibinfo{year}{1979}).

\bibitem[{\citenamefont{Gonsalves et~al.}(2019)\citenamefont{Gonsalves,
  Nakamura, Daniels, Benedetti, Pieronek, De~Raadt, Steinke, Bin, Bulanov,
  Van~Tilborg et~al.}}]{gonsalves2019petawatt}
\bibinfo{author}{\bibfnamefont{A.}~\bibnamefont{Gonsalves}},
  \bibinfo{author}{\bibfnamefont{K.}~\bibnamefont{Nakamura}},
  \bibinfo{author}{\bibfnamefont{J.}~\bibnamefont{Daniels}},
  \bibinfo{author}{\bibfnamefont{C.}~\bibnamefont{Benedetti}},
  \bibinfo{author}{\bibfnamefont{C.}~\bibnamefont{Pieronek}},
  \bibinfo{author}{\bibfnamefont{T.}~\bibnamefont{De~Raadt}},
  \bibinfo{author}{\bibfnamefont{S.}~\bibnamefont{Steinke}},
  \bibinfo{author}{\bibfnamefont{J.}~\bibnamefont{Bin}},
  \bibinfo{author}{\bibfnamefont{S.}~\bibnamefont{Bulanov}},
  \bibinfo{author}{\bibfnamefont{J.}~\bibnamefont{Van~Tilborg}},
  \bibnamefont{et~al.}, \bibinfo{journal}{Physical Review Letters}
  \textbf{\bibinfo{volume}{122}}, \bibinfo{pages}{084801}
  (\bibinfo{year}{2019}).

\bibitem[{\citenamefont{Esarey et~al.}(1995)\citenamefont{Esarey, Sprangle, and
  Krall}}]{esarey1995laser}
\bibinfo{author}{\bibfnamefont{E.}~\bibnamefont{Esarey}},
  \bibinfo{author}{\bibfnamefont{P.}~\bibnamefont{Sprangle}}, \bibnamefont{and}
  \bibinfo{author}{\bibfnamefont{J.}~\bibnamefont{Krall}},
  \bibinfo{journal}{Physical Review E} \textbf{\bibinfo{volume}{52}},
  \bibinfo{pages}{5443} (\bibinfo{year}{1995}).

\bibitem[{\citenamefont{Malka et~al.}(1997)\citenamefont{Malka, Lefebvre, and
  Miquel}}]{malka1997experimental}
\bibinfo{author}{\bibfnamefont{G.}~\bibnamefont{Malka}},
  \bibinfo{author}{\bibfnamefont{E.}~\bibnamefont{Lefebvre}}, \bibnamefont{and}
  \bibinfo{author}{\bibfnamefont{J.}~\bibnamefont{Miquel}},
  \bibinfo{journal}{Physical Review Letters} \textbf{\bibinfo{volume}{78}},
  \bibinfo{pages}{3314} (\bibinfo{year}{1997}).

\bibitem[{\citenamefont{Varin and Pich{\'e}}(2006)}]{varin2006relativistic}
\bibinfo{author}{\bibfnamefont{C.}~\bibnamefont{Varin}} \bibnamefont{and}
  \bibinfo{author}{\bibfnamefont{M.}~\bibnamefont{Pich{\'e}}},
  \bibinfo{journal}{Physical Review E} \textbf{\bibinfo{volume}{74}},
  \bibinfo{pages}{045602} (\bibinfo{year}{2006}).

\bibitem[{\citenamefont{De~Andres et~al.}(2024)\citenamefont{De~Andres,
  Bhadoria, Marmolejo, Muschet, Fischer, Reza~Barzegar, Blackburn, Gonoskov,
  Hanstorp, Marklund et~al.}}]{de2024unforeseen}
\bibinfo{author}{\bibfnamefont{A.}~\bibnamefont{De~Andres}},
  \bibinfo{author}{\bibfnamefont{S.}~\bibnamefont{Bhadoria}},
  \bibinfo{author}{\bibfnamefont{J.~T.} \bibnamefont{Marmolejo}},
  \bibinfo{author}{\bibfnamefont{A.}~\bibnamefont{Muschet}},
  \bibinfo{author}{\bibfnamefont{P.}~\bibnamefont{Fischer}},
  \bibinfo{author}{\bibfnamefont{H.}~\bibnamefont{Reza~Barzegar}},
  \bibinfo{author}{\bibfnamefont{T.}~\bibnamefont{Blackburn}},
  \bibinfo{author}{\bibfnamefont{A.}~\bibnamefont{Gonoskov}},
  \bibinfo{author}{\bibfnamefont{D.}~\bibnamefont{Hanstorp}},
  \bibinfo{author}{\bibfnamefont{M.}~\bibnamefont{Marklund}},
  \bibnamefont{et~al.}, \bibinfo{journal}{Communications Physics}
  \textbf{\bibinfo{volume}{7}}, \bibinfo{pages}{293} (\bibinfo{year}{2024}).

\bibitem[{\citenamefont{Th{\'e}venet et~al.}(2016)\citenamefont{Th{\'e}venet,
  Leblanc, Kahaly, Vincenti, Vernier, Qu{\'e}r{\'e}, and
  Faure}}]{thevenet2016vacuum}
\bibinfo{author}{\bibfnamefont{M.}~\bibnamefont{Th{\'e}venet}},
  \bibinfo{author}{\bibfnamefont{A.}~\bibnamefont{Leblanc}},
  \bibinfo{author}{\bibfnamefont{S.}~\bibnamefont{Kahaly}},
  \bibinfo{author}{\bibfnamefont{H.}~\bibnamefont{Vincenti}},
  \bibinfo{author}{\bibfnamefont{A.}~\bibnamefont{Vernier}},
  \bibinfo{author}{\bibfnamefont{F.}~\bibnamefont{Qu{\'e}r{\'e}}},
  \bibnamefont{and} \bibinfo{author}{\bibfnamefont{J.}~\bibnamefont{Faure}},
  \bibinfo{journal}{Nature Physics} \textbf{\bibinfo{volume}{12}},
  \bibinfo{pages}{355} (\bibinfo{year}{2016}).

\bibitem[{\citenamefont{Bulanov et~al.}(1994)\citenamefont{Bulanov, Naumova,
  and Pegoraro}}]{bulanov1994interaction}
\bibinfo{author}{\bibfnamefont{S.~V.} \bibnamefont{Bulanov}},
  \bibinfo{author}{\bibfnamefont{N.}~\bibnamefont{Naumova}}, \bibnamefont{and}
  \bibinfo{author}{\bibfnamefont{F.}~\bibnamefont{Pegoraro}},
  \bibinfo{journal}{Physics of Plasmas} \textbf{\bibinfo{volume}{1}},
  \bibinfo{pages}{745} (\bibinfo{year}{1994}).

\bibitem[{\citenamefont{Lichters et~al.}(1996)\citenamefont{Lichters, Meyer-ter
  Vehn, and Pukhov}}]{lichters1996short}
\bibinfo{author}{\bibfnamefont{R.}~\bibnamefont{Lichters}},
  \bibinfo{author}{\bibfnamefont{J.}~\bibnamefont{Meyer-ter Vehn}},
  \bibnamefont{and} \bibinfo{author}{\bibfnamefont{A.}~\bibnamefont{Pukhov}},
  \bibinfo{journal}{Physics of Plasmas} \textbf{\bibinfo{volume}{3}},
  \bibinfo{pages}{3425} (\bibinfo{year}{1996}).

\bibitem[{\citenamefont{Baeva et~al.}(2006)\citenamefont{Baeva, Gordienko, and
  Pukhov}}]{baeva2006theory}
\bibinfo{author}{\bibfnamefont{T.}~\bibnamefont{Baeva}},
  \bibinfo{author}{\bibfnamefont{S.}~\bibnamefont{Gordienko}},
  \bibnamefont{and} \bibinfo{author}{\bibfnamefont{A.}~\bibnamefont{Pukhov}},
  \bibinfo{journal}{Physical Review E} \textbf{\bibinfo{volume}{74}},
  \bibinfo{pages}{046404} (\bibinfo{year}{2006}).

\bibitem[{\citenamefont{Thaury et~al.}(2007)\citenamefont{Thaury, Quere,
  Geindre, Levy, Ceccotti, Monot, Bougeard, R{\'e}au, d’Oliveira, Audebert
  et~al.}}]{thaury2007plasma}
\bibinfo{author}{\bibfnamefont{C.}~\bibnamefont{Thaury}},
  \bibinfo{author}{\bibfnamefont{F.}~\bibnamefont{Quere}},
  \bibinfo{author}{\bibfnamefont{J.-P.} \bibnamefont{Geindre}},
  \bibinfo{author}{\bibfnamefont{A.}~\bibnamefont{Levy}},
  \bibinfo{author}{\bibfnamefont{T.}~\bibnamefont{Ceccotti}},
  \bibinfo{author}{\bibfnamefont{P.}~\bibnamefont{Monot}},
  \bibinfo{author}{\bibfnamefont{M.}~\bibnamefont{Bougeard}},
  \bibinfo{author}{\bibfnamefont{F.}~\bibnamefont{R{\'e}au}},
  \bibinfo{author}{\bibfnamefont{P.}~\bibnamefont{d’Oliveira}},
  \bibinfo{author}{\bibfnamefont{P.}~\bibnamefont{Audebert}},
  \bibnamefont{et~al.}, \bibinfo{journal}{Nature Physics}
  \textbf{\bibinfo{volume}{3}}, \bibinfo{pages}{424} (\bibinfo{year}{2007}).

\bibitem[{\citenamefont{Za\"{\i}m et~al.}(2017)\citenamefont{Za\"{\i}m,
  Th\'evenet, Lifschitz, and Faure}}]{PhysRevLett.119.094801}
\bibinfo{author}{\bibfnamefont{N.}~\bibnamefont{Za\"{\i}m}},
  \bibinfo{author}{\bibfnamefont{M.}~\bibnamefont{Th\'evenet}},
  \bibinfo{author}{\bibfnamefont{A.}~\bibnamefont{Lifschitz}},
  \bibnamefont{and} \bibinfo{author}{\bibfnamefont{J.}~\bibnamefont{Faure}},
  \bibinfo{journal}{Phys. Rev. Lett.} \textbf{\bibinfo{volume}{119}},
  \bibinfo{pages}{094801} (\bibinfo{year}{2017}),
  \urlprefix\url{https://link.aps.org/doi/10.1103/PhysRevLett.119.094801}.

\bibitem[{\citenamefont{Za{\"\i}m et~al.}(2020)\citenamefont{Za{\"\i}m,
  Gu{\'e}not, Chopineau, Denoeud, Lundh, Vincenti, Qu{\'e}r{\'e}, and
  Faure}}]{zaim2020interaction}
\bibinfo{author}{\bibfnamefont{N.}~\bibnamefont{Za{\"\i}m}},
  \bibinfo{author}{\bibfnamefont{D.}~\bibnamefont{Gu{\'e}not}},
  \bibinfo{author}{\bibfnamefont{L.}~\bibnamefont{Chopineau}},
  \bibinfo{author}{\bibfnamefont{A.}~\bibnamefont{Denoeud}},
  \bibinfo{author}{\bibfnamefont{O.}~\bibnamefont{Lundh}},
  \bibinfo{author}{\bibfnamefont{H.}~\bibnamefont{Vincenti}},
  \bibinfo{author}{\bibfnamefont{F.}~\bibnamefont{Qu{\'e}r{\'e}}},
  \bibnamefont{and} \bibinfo{author}{\bibfnamefont{J.}~\bibnamefont{Faure}},
  \bibinfo{journal}{Physical Review X} \textbf{\bibinfo{volume}{10}},
  \bibinfo{pages}{041064} (\bibinfo{year}{2020}).

\bibitem[{\citenamefont{Cao et~al.}(2021)\citenamefont{Cao, Hu, Hu, Zhao, Zou,
  Yang, Zhang, Shao, and Yu}}]{cao2021direct}
\bibinfo{author}{\bibfnamefont{Y.}~\bibnamefont{Cao}},
  \bibinfo{author}{\bibfnamefont{L.-X.} \bibnamefont{Hu}},
  \bibinfo{author}{\bibfnamefont{Y.}~\bibnamefont{Hu}},
  \bibinfo{author}{\bibfnamefont{J.}~\bibnamefont{Zhao}},
  \bibinfo{author}{\bibfnamefont{D.}~\bibnamefont{Zou}},
  \bibinfo{author}{\bibfnamefont{X.}~\bibnamefont{Yang}},
  \bibinfo{author}{\bibfnamefont{F.}~\bibnamefont{Zhang}},
  \bibinfo{author}{\bibfnamefont{F.}~\bibnamefont{Shao}}, \bibnamefont{and}
  \bibinfo{author}{\bibfnamefont{T.-P.} \bibnamefont{Yu}},
  \bibinfo{journal}{Optics Express} \textbf{\bibinfo{volume}{29}},
  \bibinfo{pages}{30223} (\bibinfo{year}{2021}).

\bibitem[{\citenamefont{Shi et~al.}(2021)\citenamefont{Shi, Blackman, Stutman,
  and Arefiev}}]{shi2021generation}
\bibinfo{author}{\bibfnamefont{Y.}~\bibnamefont{Shi}},
  \bibinfo{author}{\bibfnamefont{D.}~\bibnamefont{Blackman}},
  \bibinfo{author}{\bibfnamefont{D.}~\bibnamefont{Stutman}}, \bibnamefont{and}
  \bibinfo{author}{\bibfnamefont{A.}~\bibnamefont{Arefiev}},
  \bibinfo{journal}{Physical Review Letters} \textbf{\bibinfo{volume}{126}},
  \bibinfo{pages}{234801} (\bibinfo{year}{2021}).

\bibitem[{\citenamefont{Froula et~al.}(2019)\citenamefont{Froula, Palastro,
  Turnbull, Davies, Nguyen, Howard, Ramsey, Franke, Bahk, Begishev
  et~al.}}]{froula2019flying}
\bibinfo{author}{\bibfnamefont{D.}~\bibnamefont{Froula}},
  \bibinfo{author}{\bibfnamefont{J.}~\bibnamefont{Palastro}},
  \bibinfo{author}{\bibfnamefont{D.}~\bibnamefont{Turnbull}},
  \bibinfo{author}{\bibfnamefont{A.}~\bibnamefont{Davies}},
  \bibinfo{author}{\bibfnamefont{L.}~\bibnamefont{Nguyen}},
  \bibinfo{author}{\bibfnamefont{A.}~\bibnamefont{Howard}},
  \bibinfo{author}{\bibfnamefont{D.}~\bibnamefont{Ramsey}},
  \bibinfo{author}{\bibfnamefont{P.}~\bibnamefont{Franke}},
  \bibinfo{author}{\bibfnamefont{S.-W.} \bibnamefont{Bahk}},
  \bibinfo{author}{\bibfnamefont{I.}~\bibnamefont{Begishev}},
  \bibnamefont{et~al.}, \bibinfo{journal}{Physics of Plasmas}
  \textbf{\bibinfo{volume}{26}}, \bibinfo{pages}{032109}
  (\bibinfo{year}{2019}).

\bibitem[{\citenamefont{Jolly et~al.}(2020)\citenamefont{Jolly, Gobert,
  Jeandet, and Qu{\'e}r{\'e}}}]{jolly2020controlling}
\bibinfo{author}{\bibfnamefont{S.~W.} \bibnamefont{Jolly}},
  \bibinfo{author}{\bibfnamefont{O.}~\bibnamefont{Gobert}},
  \bibinfo{author}{\bibfnamefont{A.}~\bibnamefont{Jeandet}}, \bibnamefont{and}
  \bibinfo{author}{\bibfnamefont{F.}~\bibnamefont{Qu{\'e}r{\'e}}},
  \bibinfo{journal}{Optics Express} \textbf{\bibinfo{volume}{28}},
  \bibinfo{pages}{4888} (\bibinfo{year}{2020}).

\bibitem[{\citenamefont{Froula et~al.}(2018)\citenamefont{Froula, Turnbull,
  Davies, Kessler, Haberberger, Palastro, Bahk, Begishev, Boni, Bucht
  et~al.}}]{froula2018spatiotemporal}
\bibinfo{author}{\bibfnamefont{D.~H.} \bibnamefont{Froula}},
  \bibinfo{author}{\bibfnamefont{D.}~\bibnamefont{Turnbull}},
  \bibinfo{author}{\bibfnamefont{A.~S.} \bibnamefont{Davies}},
  \bibinfo{author}{\bibfnamefont{T.~J.} \bibnamefont{Kessler}},
  \bibinfo{author}{\bibfnamefont{D.}~\bibnamefont{Haberberger}},
  \bibinfo{author}{\bibfnamefont{J.~P.} \bibnamefont{Palastro}},
  \bibinfo{author}{\bibfnamefont{S.-W.} \bibnamefont{Bahk}},
  \bibinfo{author}{\bibfnamefont{I.~A.} \bibnamefont{Begishev}},
  \bibinfo{author}{\bibfnamefont{R.}~\bibnamefont{Boni}},
  \bibinfo{author}{\bibfnamefont{S.}~\bibnamefont{Bucht}},
  \bibnamefont{et~al.}, \bibinfo{journal}{Nature Photonics}
  \textbf{\bibinfo{volume}{12}}, \bibinfo{pages}{262} (\bibinfo{year}{2018}).

\bibitem[{\citenamefont{Sainte-Marie et~al.}(2017)\citenamefont{Sainte-Marie,
  Gobert, and Quere}}]{sainte2017controlling}
\bibinfo{author}{\bibfnamefont{A.}~\bibnamefont{Sainte-Marie}},
  \bibinfo{author}{\bibfnamefont{O.}~\bibnamefont{Gobert}}, \bibnamefont{and}
  \bibinfo{author}{\bibfnamefont{F.}~\bibnamefont{Quere}},
  \bibinfo{journal}{Optica} \textbf{\bibinfo{volume}{4}}, \bibinfo{pages}{1298}
  (\bibinfo{year}{2017}).

\bibitem[{\citenamefont{Smartsev et~al.}(2019)\citenamefont{Smartsev,
  Caizergues, Oubrerie, Gautier, Goddet, Tafzi, Phuoc, Malka, and
  Thaury}}]{smartsev2019axiparabola}
\bibinfo{author}{\bibfnamefont{S.}~\bibnamefont{Smartsev}},
  \bibinfo{author}{\bibfnamefont{C.}~\bibnamefont{Caizergues}},
  \bibinfo{author}{\bibfnamefont{K.}~\bibnamefont{Oubrerie}},
  \bibinfo{author}{\bibfnamefont{J.}~\bibnamefont{Gautier}},
  \bibinfo{author}{\bibfnamefont{J.-P.} \bibnamefont{Goddet}},
  \bibinfo{author}{\bibfnamefont{A.}~\bibnamefont{Tafzi}},
  \bibinfo{author}{\bibfnamefont{K.~T.} \bibnamefont{Phuoc}},
  \bibinfo{author}{\bibfnamefont{V.}~\bibnamefont{Malka}}, \bibnamefont{and}
  \bibinfo{author}{\bibfnamefont{C.}~\bibnamefont{Thaury}},
  \bibinfo{journal}{Optics Letters} \textbf{\bibinfo{volume}{44}},
  \bibinfo{pages}{3414} (\bibinfo{year}{2019}).

\bibitem[{\citenamefont{Simpson et~al.}(2022)\citenamefont{Simpson, Ramsey,
  Franke, Weichman, Ambat, Turnbull, Froula, and
  Palastro}}]{simpson2022spatiotemporal}
\bibinfo{author}{\bibfnamefont{T.~T.} \bibnamefont{Simpson}},
  \bibinfo{author}{\bibfnamefont{D.}~\bibnamefont{Ramsey}},
  \bibinfo{author}{\bibfnamefont{P.}~\bibnamefont{Franke}},
  \bibinfo{author}{\bibfnamefont{K.}~\bibnamefont{Weichman}},
  \bibinfo{author}{\bibfnamefont{M.~V.} \bibnamefont{Ambat}},
  \bibinfo{author}{\bibfnamefont{D.}~\bibnamefont{Turnbull}},
  \bibinfo{author}{\bibfnamefont{D.~H.} \bibnamefont{Froula}},
  \bibnamefont{and} \bibinfo{author}{\bibfnamefont{J.~P.}
  \bibnamefont{Palastro}}, \bibinfo{journal}{Optics Express}
  \textbf{\bibinfo{volume}{30}}, \bibinfo{pages}{9878} (\bibinfo{year}{2022}).

\bibitem[{\citenamefont{Ambat et~al.}(2023)\citenamefont{Ambat, Shaw, Pigeon,
  Miller, Simpson, Froula, and Palastro}}]{ambat2023programmable}
\bibinfo{author}{\bibfnamefont{M.}~\bibnamefont{Ambat}},
  \bibinfo{author}{\bibfnamefont{J.}~\bibnamefont{Shaw}},
  \bibinfo{author}{\bibfnamefont{J.}~\bibnamefont{Pigeon}},
  \bibinfo{author}{\bibfnamefont{K.}~\bibnamefont{Miller}},
  \bibinfo{author}{\bibfnamefont{T.}~\bibnamefont{Simpson}},
  \bibinfo{author}{\bibfnamefont{D.}~\bibnamefont{Froula}}, \bibnamefont{and}
  \bibinfo{author}{\bibfnamefont{J.~P.} \bibnamefont{Palastro}},
  \bibinfo{journal}{Optics Express} \textbf{\bibinfo{volume}{31}},
  \bibinfo{pages}{31354} (\bibinfo{year}{2023}).

\bibitem[{\citenamefont{Pigeon et~al.}(2023)\citenamefont{Pigeon, Franke, Lim
  Pac~Chong, Katz, Boni, Dorrer, Palastro, and
  Froula}}]{pigeon2023ultrabroadband}
\bibinfo{author}{\bibfnamefont{J.}~\bibnamefont{Pigeon}},
  \bibinfo{author}{\bibfnamefont{P.}~\bibnamefont{Franke}},
  \bibinfo{author}{\bibfnamefont{M.}~\bibnamefont{Lim Pac~Chong}},
  \bibinfo{author}{\bibfnamefont{J.}~\bibnamefont{Katz}},
  \bibinfo{author}{\bibfnamefont{R.}~\bibnamefont{Boni}},
  \bibinfo{author}{\bibfnamefont{C.}~\bibnamefont{Dorrer}},
  \bibinfo{author}{\bibfnamefont{J.~P.} \bibnamefont{Palastro}},
  \bibnamefont{and} \bibinfo{author}{\bibfnamefont{D.}~\bibnamefont{Froula}},
  \bibinfo{journal}{Optics Express} \textbf{\bibinfo{volume}{32}},
  \bibinfo{pages}{576} (\bibinfo{year}{2023}).

\bibitem[{\citenamefont{Pierce et~al.}(2023)\citenamefont{Pierce, Palastro, Li,
  Malaca, Ramsey, Vieira, Weichman, and Mori}}]{PhysRevResearch.5.013085}
\bibinfo{author}{\bibfnamefont{J.~R.} \bibnamefont{Pierce}},
  \bibinfo{author}{\bibfnamefont{J.~P.} \bibnamefont{Palastro}},
  \bibinfo{author}{\bibfnamefont{F.}~\bibnamefont{Li}},
  \bibinfo{author}{\bibfnamefont{B.}~\bibnamefont{Malaca}},
  \bibinfo{author}{\bibfnamefont{D.}~\bibnamefont{Ramsey}},
  \bibinfo{author}{\bibfnamefont{J.}~\bibnamefont{Vieira}},
  \bibinfo{author}{\bibfnamefont{K.}~\bibnamefont{Weichman}}, \bibnamefont{and}
  \bibinfo{author}{\bibfnamefont{W.~B.} \bibnamefont{Mori}},
  \bibinfo{journal}{Phys. Rev. Res.} \textbf{\bibinfo{volume}{5}},
  \bibinfo{pages}{013085} (\bibinfo{year}{2023}),
  \urlprefix\url{https://link.aps.org/doi/10.1103/PhysRevResearch.5.013085}.

\bibitem[{\citenamefont{Palastro et~al.}(2020)\citenamefont{Palastro, Shaw,
  Franke, Ramsey, Simpson, and Froula}}]{palastro_dephasingless_2020}
\bibinfo{author}{\bibfnamefont{J.}~\bibnamefont{Palastro}},
  \bibinfo{author}{\bibfnamefont{J.~L.} \bibnamefont{Shaw}},
  \bibinfo{author}{\bibfnamefont{P.}~\bibnamefont{Franke}},
  \bibinfo{author}{\bibfnamefont{D.}~\bibnamefont{Ramsey}},
  \bibinfo{author}{\bibfnamefont{T.~T.} \bibnamefont{Simpson}},
  \bibnamefont{and} \bibinfo{author}{\bibfnamefont{D.~H.}
  \bibnamefont{Froula}}, \bibinfo{journal}{Phys. Rev. Lett.}
  \textbf{\bibinfo{volume}{124}}, \bibinfo{pages}{134802}
  (\bibinfo{year}{2020}).

\bibitem[{\citenamefont{Caizergues et~al.}(2020)\citenamefont{Caizergues,
  Smartsev, Malka, and Thaury}}]{caizergues_phase-locked_2020}
\bibinfo{author}{\bibfnamefont{C.}~\bibnamefont{Caizergues}},
  \bibinfo{author}{\bibfnamefont{S.}~\bibnamefont{Smartsev}},
  \bibinfo{author}{\bibfnamefont{V.}~\bibnamefont{Malka}}, \bibnamefont{and}
  \bibinfo{author}{\bibfnamefont{C.}~\bibnamefont{Thaury}},
  \bibinfo{journal}{Nat. Photonics} \textbf{\bibinfo{volume}{14}},
  \bibinfo{pages}{475} (\bibinfo{year}{2020}), ISSN \bibinfo{issn}{1749-4893},
  \urlprefix\url{https://www.nature.com/articles/s41566-020-0657-2}.

\bibitem[{\citenamefont{Formanek et~al.}(2023)\citenamefont{Formanek, Palastro,
  Vranic, Ramsey, and Di~Piazza}}]{formanek_charged_2023}
\bibinfo{author}{\bibfnamefont{M.}~\bibnamefont{Formanek}},
  \bibinfo{author}{\bibfnamefont{J.~P.} \bibnamefont{Palastro}},
  \bibinfo{author}{\bibfnamefont{M.}~\bibnamefont{Vranic}},
  \bibinfo{author}{\bibfnamefont{D.}~\bibnamefont{Ramsey}}, \bibnamefont{and}
  \bibinfo{author}{\bibfnamefont{A.}~\bibnamefont{Di~Piazza}},
  \bibinfo{journal}{Phys. Rev. E} \textbf{\bibinfo{volume}{107}},
  \bibinfo{pages}{055213} (\bibinfo{year}{2023}).

\bibitem[{\citenamefont{Ramsey et~al.}(2020)\citenamefont{Ramsey, Franke,
  Simpson, Froula, and Palastro}}]{ramsey2020vacuum}
\bibinfo{author}{\bibfnamefont{D.}~\bibnamefont{Ramsey}},
  \bibinfo{author}{\bibfnamefont{P.}~\bibnamefont{Franke}},
  \bibinfo{author}{\bibfnamefont{T.}~\bibnamefont{Simpson}},
  \bibinfo{author}{\bibfnamefont{D.}~\bibnamefont{Froula}}, \bibnamefont{and}
  \bibinfo{author}{\bibfnamefont{J.}~\bibnamefont{Palastro}},
  \bibinfo{journal}{Physical Review E} \textbf{\bibinfo{volume}{102}},
  \bibinfo{pages}{043207} (\bibinfo{year}{2020}).

\bibitem[{\citenamefont{Ramsey et~al.}(2022)\citenamefont{Ramsey, Malaca,
  Di~Piazza, Formanek, Franke, Froula, Pardal, Simpson, Vieira, Weichman
  et~al.}}]{ramsey2022nonlinear}
\bibinfo{author}{\bibfnamefont{D.}~\bibnamefont{Ramsey}},
  \bibinfo{author}{\bibfnamefont{B.}~\bibnamefont{Malaca}},
  \bibinfo{author}{\bibfnamefont{A.}~\bibnamefont{Di~Piazza}},
  \bibinfo{author}{\bibfnamefont{M.}~\bibnamefont{Formanek}},
  \bibinfo{author}{\bibfnamefont{P.}~\bibnamefont{Franke}},
  \bibinfo{author}{\bibfnamefont{D.}~\bibnamefont{Froula}},
  \bibinfo{author}{\bibfnamefont{M.}~\bibnamefont{Pardal}},
  \bibinfo{author}{\bibfnamefont{T.}~\bibnamefont{Simpson}},
  \bibinfo{author}{\bibfnamefont{J.}~\bibnamefont{Vieira}},
  \bibinfo{author}{\bibfnamefont{K.}~\bibnamefont{Weichman}},
  \bibnamefont{et~al.}, \bibinfo{journal}{Physical Review E}
  \textbf{\bibinfo{volume}{105}}, \bibinfo{pages}{065201}
  (\bibinfo{year}{2022}).

\bibitem[{\citenamefont{Pukhov}(1999)}]{pukhov1999three}
\bibinfo{author}{\bibfnamefont{A.}~\bibnamefont{Pukhov}},
  \bibinfo{journal}{Journal of Plasma Physics} \textbf{\bibinfo{volume}{61}},
  \bibinfo{pages}{425} (\bibinfo{year}{1999}).

\bibitem[{\citenamefont{Ramsey et~al.}(2023)\citenamefont{Ramsey, Di~Piazza,
  Formanek, Franke, Froula, Malaca, Mori, Pierce, Simpson, Vieira
  et~al.}}]{ramsey2023exact}
\bibinfo{author}{\bibfnamefont{D.}~\bibnamefont{Ramsey}},
  \bibinfo{author}{\bibfnamefont{A.}~\bibnamefont{Di~Piazza}},
  \bibinfo{author}{\bibfnamefont{M.}~\bibnamefont{Formanek}},
  \bibinfo{author}{\bibfnamefont{P.}~\bibnamefont{Franke}},
  \bibinfo{author}{\bibfnamefont{D.}~\bibnamefont{Froula}},
  \bibinfo{author}{\bibfnamefont{B.}~\bibnamefont{Malaca}},
  \bibinfo{author}{\bibfnamefont{W.}~\bibnamefont{Mori}},
  \bibinfo{author}{\bibfnamefont{J.}~\bibnamefont{Pierce}},
  \bibinfo{author}{\bibfnamefont{T.}~\bibnamefont{Simpson}},
  \bibinfo{author}{\bibfnamefont{J.}~\bibnamefont{Vieira}},
  \bibnamefont{et~al.}, \bibinfo{journal}{Physical Review A}
  \textbf{\bibinfo{volume}{107}}, \bibinfo{pages}{013513}
  (\bibinfo{year}{2023}).

\bibitem[{\citenamefont{Franke et~al.}(2021)\citenamefont{Franke, Ramsey,
  Simpson, Turnbull, Froula, and Palastro}}]{franke_optical_2021}
\bibinfo{author}{\bibfnamefont{P.}~\bibnamefont{Franke}},
  \bibinfo{author}{\bibfnamefont{D.}~\bibnamefont{Ramsey}},
  \bibinfo{author}{\bibfnamefont{T.~T.} \bibnamefont{Simpson}},
  \bibinfo{author}{\bibfnamefont{D.}~\bibnamefont{Turnbull}},
  \bibinfo{author}{\bibfnamefont{D.~H.} \bibnamefont{Froula}},
  \bibnamefont{and} \bibinfo{author}{\bibfnamefont{J.~P.}
  \bibnamefont{Palastro}}, \bibinfo{journal}{Phys. Rev. A}
  \textbf{\bibinfo{volume}{104}}, \bibinfo{pages}{043520}
  (\bibinfo{year}{2021}).

\bibitem[{\citenamefont{Courtois et~al.}(2011)\citenamefont{Courtois, Edwards,
  Compant La~Fontaine, Aedy, Barbotin, Bazzoli, Biddle, Brebion, Bourgade, Drew
  et~al.}}]{courtois2011high}
\bibinfo{author}{\bibfnamefont{C.}~\bibnamefont{Courtois}},
  \bibinfo{author}{\bibfnamefont{R.}~\bibnamefont{Edwards}},
  \bibinfo{author}{\bibfnamefont{A.}~\bibnamefont{Compant La~Fontaine}},
  \bibinfo{author}{\bibfnamefont{C.}~\bibnamefont{Aedy}},
  \bibinfo{author}{\bibfnamefont{M.}~\bibnamefont{Barbotin}},
  \bibinfo{author}{\bibfnamefont{S.}~\bibnamefont{Bazzoli}},
  \bibinfo{author}{\bibfnamefont{L.}~\bibnamefont{Biddle}},
  \bibinfo{author}{\bibfnamefont{D.}~\bibnamefont{Brebion}},
  \bibinfo{author}{\bibfnamefont{J.}~\bibnamefont{Bourgade}},
  \bibinfo{author}{\bibfnamefont{D.}~\bibnamefont{Drew}}, \bibnamefont{et~al.},
  \bibinfo{journal}{Physics of Plasmas} \textbf{\bibinfo{volume}{18}},
  \bibinfo{pages}{023101} (\bibinfo{year}{2011}).

\bibitem[{\citenamefont{Bruhaug et~al.}(2023)\citenamefont{Bruhaug, Freeman,
  Rinderknecht, Neukirch, Wilde, Merrill, Rygg, Wei, Collins, and
  Shaw}}]{bruhaug2023single}
\bibinfo{author}{\bibfnamefont{G.}~\bibnamefont{Bruhaug}},
  \bibinfo{author}{\bibfnamefont{M.}~\bibnamefont{Freeman}},
  \bibinfo{author}{\bibfnamefont{H.}~\bibnamefont{Rinderknecht}},
  \bibinfo{author}{\bibfnamefont{L.}~\bibnamefont{Neukirch}},
  \bibinfo{author}{\bibfnamefont{C.}~\bibnamefont{Wilde}},
  \bibinfo{author}{\bibfnamefont{F.}~\bibnamefont{Merrill}},
  \bibinfo{author}{\bibfnamefont{J.}~\bibnamefont{Rygg}},
  \bibinfo{author}{\bibfnamefont{M.}~\bibnamefont{Wei}},
  \bibinfo{author}{\bibfnamefont{G.}~\bibnamefont{Collins}}, \bibnamefont{and}
  \bibinfo{author}{\bibfnamefont{J.}~\bibnamefont{Shaw}},
  \bibinfo{journal}{Sci. Rep.} \textbf{\bibinfo{volume}{13}},
  \bibinfo{pages}{2227} (\bibinfo{year}{2023}).

\end{thebibliography}
\end{document}